\documentclass[twocolumn]{revtex4}
\usepackage[dvips]{graphicx}
\usepackage{amsfonts}
\usepackage{amsmath}
\usepackage{float, color,array, multirow}

\begin{document}

\title{RKKY interaction in the presence of a charge-density-wave order}

\author{Kazushi Aoyama}

\date{\today}

\affiliation{Department of Earth and Space Science, Graduate School of Science, Osaka University, Osaka 560-0043, Japan}

\begin{abstract}
Effects of a charge-density-wave (CDW) order on the Ruderman–Kittel–Kasuya–Yosida (RKKY) interaction has been theoretically investigated. Assuming that the CDW with an incommensurate ordering vector ${\bf Q}_c$ is induced by a Fermi surface nesting, we show that the CDW order suppresses the conventional RKKY interaction and that it can induce a mode coupling between magnetic ordering vectors ${\bf q}_m$ and ${\bf q}_m\pm {\bf Q}_c$. This suggests that below the CDW transition temperature, the correlation between localized spins may develop at both ${\bf q}_m$ and ${\bf q}_m\pm {\bf Q}_c$. Experimental implications of our result to van der Waals materials $R$Te$_3$ are also discussed. 
\end{abstract}

\maketitle
Competition and coexistence of ordered phases have attracted much attention in condensed matter physics, and superconductivity influenced by magnetic, charge, and orbital orderings has extensively been studied so far \cite{HF_Steglich_jpsj_05, Cuprate_Lee_rmp_06, FeSC_Stewart_rmp_11, OrganicSC_Ardavan_jpsj_12}. Concerning an interplay between magnetic and charge orderings in metallic systems, however, less is know about it particularly in the absence of spin-orbit couplings. In this work, we theoretically investigate an interaction between localized spins ${\bf S}_i$'s mediated by conduction electrons, i.e., the Ruderman–Kittel–Kasuya–Yosida (RKKY) interaction, in a situation where the electrons are condensed into a charge-density-wave (CDW) state.     

\begin{figure}[t]
\includegraphics[scale = 0.5]{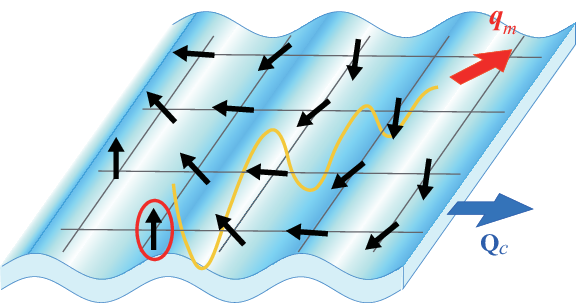}
\caption{ Schematic image of the system: spins (black arrows) on a lattice in the presence of a CDW order (blue region), where CDW and magnetic ordering-wave-vectors, ${\bf Q}_c$ and ${\bf q}_m$, are indicated by blue and red arrows, respectively. A yellow wavy curve represents the conventional RKKY interaction with its origin at the spin enclosed by a red circle. \label{fig:fig0}}
\end{figure}

In magnetic metals where localized and conduction-electron spins are interacting with each other via the Kondo coupling of the form ${\cal H}_{K}= J \sum_{i,s,s'} {\bf S}_i \cdot{\mbox {\boldmath $\sigma$}}_{ss'} \hat{c}_{i,s}^\dagger  \hat{c}_{i,s'}$ with $i$-site creation (annihilation) operator of the conduction electron with spin $s$ $\hat{c}^\dagger_{i ,s}$ ($\hat{c}_{i ,s}$) and Pauli matrices ${\mbox {\boldmath $\sigma$}}=(\sigma_x,\sigma_y,\sigma_z)$, magnetic properties are basically governed by the localized spins ${\bf S}_i$'s which are correlated to each other over a long distance as the conduction electrons can carry the spin information. Such a conduction-electron mediated effective interaction between ${\bf S}_i$'s is called the RKKY interaction. Since the RKKY interaction exhibits a damping oscillation as a function of distance (see a yellow wavy line in Fig. \ref{fig:fig0}) and its oscillation period is determined by the Fermi wave number $k_F$, the ordering vector ${\bf q}_m$ of the localized spin ${\bf S}_i$ generally depends on the details of lattice and electronic structures \cite{2DRKKY_Wang_prl_20, 3DRKKY_Mitsumoto_prb_21}. In the conventional theory of the RKKY interaction, it is assumed that the conduction electrons are in the normal metallic state. Then, the question is how the RKKY interaction is modified when the electrons undergo phase transitions into long-range ordered states, one interesting example of which is a CDW state possessing a charge ordering vector ${\bf Q}_c$ (see Fig. \ref{fig:fig0}).  

The CDW is a periodic modulation of electron density which has mostly been observed in low-dimensional materials such as the quasi-1D conductors NbSe$_3$ and TaS$_3$ \cite{C0018} and the quasi-2D  transition-metal dichalcogenides NbSe$_2$ and TaS$_2$ \cite{2Ddichalcoge_review}. The CDW is usually caused by a Fermi-surface nesting and its period, or equivalently, the wave number of the modulation ${\bf Q}_c$, is associated with the nesting vector. So far, various aspects of CDW physics including a sliding dynamics \cite{C0018, C0070, C0014, Et-Shapiro_Thorne_prb_87, Matsukawa_JJAP_1987, FA_shapiro,FA_frqmix}, spatial patterns \cite{2DCDW_Nakanishi_jpsj_82, 2DCDW_Gye_prl_19, 2DCDW_yoshizawa_prl_24}, and excitations \cite{LF-phason_Lee_prb_78, LF-phason_Kim_natmat_23} have been studied, and of recent particular interest could be the coexistence of a magnetic order and the CDW order reported in Er$_5$Ir$_4$Si$_{10}$ \cite{Er5Ir4Si10_Galli_prl_00}, $R$NiC$_2$ ($R$=Gd, Tb) \cite{RNiC2_Shimomura_prb_16, RNiC2_Hanasaki_prb_17}, EuAl$_4$ \cite{EuAl4_Shimomura_jpsj_19, EuAl4_Kaneko_jpsj_21, EuAl4_Takagki_natcom_22, EuAl4_Gen_prb_24}, and van der Waals materials $R$Te$_3$ ($R$=Ce, Gd, Tb, Dy) \cite{CeTe3_Okuma_screp_20, GdTe3_Raghavan_npjq_24, TbTe3_Pfuner_jpcm_12, TbTe3_Chillal_prb_20, RTe3_Higashihara_prb_24, DyTe3_Akatsuka_natcom_24}. In these compounds, a magnetic transition occurs at a temperature below the CDW transition temperature, so that it is naively expected that the spin ordering would be affected by the CDW order. Actually, recent experiments on $R$Te$_3$ show that localized spins at $R$ sites are ordered into a magnetic state characterized by the several ordering vectors of ${\bf q}_m$ and ${\bf q}_m \pm {\bf Q}_c$ \cite{TbTe3_Chillal_prb_20, DyTe3_Akatsuka_natcom_24,GdTe3_Raghavan_npjq_24}, but its mechanism is not yet well understood. For example, CDW-modulation effects on orbital degrees of freedom and exchange interactions have been discussed in TbTe$_3$ \cite{TbTe3_Chillal_prb_20} and DyTe$_3$ \cite{DyTe3_Akatsuka_natcom_24}, respectively, whereas in GdTe$_3$ \cite{GdTe3_Raghavan_npjq_24}, an in-plane coupling between the CDW and a ferromagnetic moment has been suggested.

In this work, to get an insight into the origin of such an exotic magnetic state, we theoretically investigate how the CDW formation affects the RKKY interaction nesting effects on the CDW-free RKKY interaction has been studied in Refs. \cite{RKKY-nesting_hayami_prb_17, Gd2PdSi3_RKKY-nesting_Dong_prl_24}).
Assuming that the CDW is caused by the Fermi-surface nesting, we will show that at least for an incommensurate ${\bf Q}_c$, a CDW-induced additional interaction connecting the magnetic ordering vectors ${\bf q}_m$ and ${\bf q}_m \pm {\bf Q}_c$ emerges.   

Since the above magnetic materials host CDWs that are basically unidirectional, we start from the mean-field Hamiltonian for such a single-${\bf Q}_c$ CDW order 
\begin{equation}\label{eq:MF}
{\cal H}_0= \sum_{{\bf k},s} \xi_{\bf k} \, \hat{c}_{{\bf k},s}^{\dag} \hat{c}_{{\bf k},s} + \frac{1}{|g|} |W_{{\bf Q}_c}|^2 + \frac{1}{2} \sum_{{\bf q}=\pm{\bf Q}_c} \sum_{{\bf k},s} W_{\bf q} \hat{c}_{{\bf k}+{\bf q},s}^{\dag} \hat{c}_{{\bf k},s} ,
\end{equation}
where $\hat{c}_{{\bf k} ,s}$ is the Fourier transform of $\hat{c}_{i ,s}$, $\xi_{\bf k}$ is a conduction-electron energy measured from the chemical potential $\mu$, and $W_{{\bf Q}_c}=W^\ast_{-{\bf Q}_c}=|W_{{\bf Q}_c}|e^{i\phi_c}$ denotes the CDW order parameter (for the derivation of Eq. (\ref{eq:MF}), see Supplemental Material \cite{supple}). Equation (\ref{eq:MF}) corresponds to a simplified single-band version of the two-band model for $R$Te$_3$ \cite{CDW_Kivelson_prb_06}. By introducing $\hat{\psi}^\dagger_{\bf k}=( \hat{c}^\dagger_{{\bf k},\uparrow},  \hat{c}^\dagger_{{\bf k},\downarrow},  \hat{c}^\dagger_{{\bf k}+{\bf Q}_c,\uparrow},  \hat{c}^\dagger_{{\bf k}+{\bf Q}_c,\downarrow} ) $, the total Hamiltonian involving the Kondo coupling ${\cal H}={\cal H}_0+{\cal H}_K$ could be rewritten in a matrix form as
\begin{eqnarray}
&& {\cal H}= \frac{1}{|g|} |W_{{\bf Q}_c}|^2 + \frac{1}{2}\sum_{ {\bf k},{\bf k}' } \hat{\psi}^\dagger_{{\bf k}'} M_{{\bf k}'{\bf k}} \hat{\psi}_{\bf k}, \\
&& M_{{\bf k}'{\bf k}} = \left( \begin{array}{cc}
\xi_{\bf k} I & W^\ast_{{\bf Q}_c} I  \nonumber\\
W_{{\bf Q}_c} I &  \xi_{{\bf k}+{\bf Q}_c} I \nonumber
 \end{array} \right) \delta_{{\bf k}^\prime, {\bf k}} 
 +  \frac{J}{2} \, \Sigma_{{\bf k}^\prime {\bf k}},  \nonumber\\
&& \Sigma_{{\bf k}^\prime {\bf k}} =  \left( \begin{array}{cc}
{\bf S}_{{\bf k}^\prime-{\bf k}}\cdot {\mbox{\boldmath $\sigma$}} & {\bf S}_{{\bf k}^\prime-{\bf k}-{\bf Q}_c}\cdot {\mbox{\boldmath $\sigma$}}   \nonumber\\
{\bf S}_{{\bf k}^\prime-{\bf k}+{\bf Q}_c }\cdot {\mbox{\boldmath $\sigma$}} & {\bf S}_{{\bf k}^\prime-{\bf k}}\cdot {\mbox{\boldmath $\sigma$}} \nonumber\\
 \end{array} \right) 
 \end{eqnarray}
 with the Fourier transform of the localized spin ${\bf S}_{\bf q}$ and the 2$\times$2 unit matrix $I$. 
 Here, we have assumed an incommensurate CDW ordering vector ${\bf Q}_c$ for which the system is translationally invariant and a jellium picture ignoring the periodic lattice potential is known to provide a good modeling \cite{comme-potential_Lee_ssc_74}.  When ${\bf Q}_c$ is commensurate, the discrete nature of the underlying lattice is not negligible and contributions from $\hat{c}^\dagger_{{\bf k}+2{\bf Q}_c,s}$, $\hat{c}^\dagger_{{\bf k}+3{\bf Q}_c,s}$, $\cdots$ become relevant. 
In the CDW sector, these commensurate contributions result in a pinning for the CDW phase $\phi_c$ \cite{comme-potential_Lee_ssc_74}, while such an effect is absent in the incommensurate case, although in reality, $\phi_c$ is subject to another pinning by impurities or defects \cite{C0018, C0070, pinning_Fukuyama_prb_78}.
Throughout this paper, we restrict ourselves to the incommensurate-${\bf Q}_c$ case without impurities for simplicity, and ignore these contributions.  
 
 By using the unitary matrix $U_{\bf k}=(\begin{array}{cc} u_{\bf k} I & -v^\ast_{\bf k} I \\ v_{\bf k} I & u_{\bf k} I \end{array})$ with $u_{\bf k}=\big[ \frac{1}{2}(1+\frac{\gamma_{\bf k}}{\sqrt{\gamma_{\bf k}^2+ |W_{{\bf Q}_c}|^2}}) \big]^{1/2} $, $v_{\bf k}=\frac{W_{{\bf Q}_c}}{|W_{{\bf Q}_c}|}\big[\frac{1}{2}(1-\frac{\gamma_{\bf k}}{\sqrt{\gamma_{\bf k}^2+ |W_{{\bf Q}_c}|^2}}) \big]^{1/2}$, and $\gamma_{\bf k}=(\xi_{\bf k}-\xi_{{\bf k}+{\bf Q}_c})/2$, we have
\begin{equation}
U^\dagger_{{\bf k}^\prime} M_{{\bf k}'{\bf k}} U_{\bf k} = \left( \begin{array}{cc} 
E_{\bf k}^+ I & 0  \nonumber\\
0 &  - E_{\bf k}^- I \nonumber
 \end{array} \right) \delta_{{\bf k}^\prime, {\bf k}} + \frac{J}{2}  U^\dagger_{{\bf k}^\prime} \Sigma_{{\bf k}'{\bf k}} U_{\bf k} , \nonumber
 \end{equation}
where $E_{\bf k}^{\pm}=\sqrt{\gamma_{\bf k}^2+|W_{{\bf Q}_c}|^2} \pm \frac{1}{2}\big(  \xi_{\bf k}+\xi_{{\bf k}+{\bf Q}_c} \big)$. Since the CDW part of the Hamiltonian has already been diagonalized, one can derive the RKKY interaction straightforwardly by expanding the free energy ${\cal F}=- T \ln {\rm tr}_{\hat{c}^\dagger, \hat{c}} \big[ \exp(-{\cal H}/T) \big]$ up to the second order in $J$. By using the Green's function ${\cal G}^{\pm}_{\varepsilon_n}({\bf k})=(i\varepsilon_n \mp E^\pm_{\bf k})^{-1}$ instead of the conventional normal-state one ${\cal G}_{\varepsilon_n}({\bf k})=(i\varepsilon_n-\xi_{\bf k})^{-1}$ with Matsubara frequency $\varepsilon_n=(2n+1)\pi T$, we obtain 
 \begin{eqnarray}\label{eq:RKKY}
 {\cal F} &=& {\cal F}_{\rm CDW} - \Big( \frac{J}{2}\Big)^2 \sum_{\bf q} \Big[ K_0({\bf q}) {\bf S}_{\bf q}\cdot {\bf S}_{-{\bf q}}  \nonumber\\
&+&  \Big\{  K^\prime_0 ({\bf q}) {\bf S}_{{\bf q}+{\bf Q}_c}\cdot {\bf S}_{-{\bf q}-{\bf Q}_c} + K_1  ({\bf q}){\bf S}_{{\bf q}-{\bf Q}_c}\cdot {\bf S}_{-{\bf q}} \nonumber\\
&& + K_2  ({\bf q}) {\bf S}_{{\bf q}-{\bf Q}_c}\cdot {\bf S}_{-{\bf q}-{\bf Q}_c} \Big\} + \Big\{ {\rm c. \, c. }\Big\} \Big], 
\end{eqnarray}
 where the coefficients are given by
\begin{eqnarray} \label{eq:coefficient}
&& K_0({\bf q}) = \frac{1}{8}\sum_{\bf k}  \Big[ 2L^{++} + ( 2g_{\bf k}g_{{\bf k}+{\bf q}}+f^\ast_{\bf k}f_{{\bf k}+{\bf q}} + f_{\bf k}f^\ast_{{\bf k}+{\bf q}}  )L^{+-} \Big], \nonumber\\
&& K_0^\prime({\bf q}) = \frac{1}{8}\sum_{\bf k}  \Big[ L^{++} - g_{\bf k}g_{{\bf k}+{\bf q}} L^{+-} + g_{\bf k} L^{--} - g_{{\bf k}+{\bf q}} L^{-+}  \Big], \nonumber\\
&& K_1({\bf q})  =  \frac{1}{4}\sum_{\bf k}  \Big[ f_{\bf k} L^{--}+f_{{\bf k}+{\bf q}} L^{-+} + (f_{\bf k}g_{{\bf k}+{\bf q}}-g_{\bf k}f_{{\bf k}+{\bf q}}) L^{+-}  \Big], \nonumber\\
&& K_2({\bf q}) = \frac{1}{8}\sum_{\bf k} f_{\bf k} f_{{\bf k}+{\bf q}} L^{+-} \nonumber 
\end{eqnarray}
with
\begin{eqnarray}
&& g_{\bf k} = \frac{\gamma_{\bf k}}{\sqrt{\gamma_{\bf k}^2 + |W_{{\bf Q}_c}|^2}}, \quad f_{\bf k} = \frac{W_{{\bf Q}_c}}{\sqrt{\gamma_{\bf k}^2 + |W_{{\bf Q}_c}|^2}} , \nonumber\\
&& L^{s s^\prime} = F_{{\bf k},{\bf q}}^{++}+ s \, F_{{\bf k},{\bf q}}^{--}+ s^\prime \big( F_{{\bf k},{\bf q}}^{-+}+s \, F_{{\bf k},{\bf q}}^{+-} \big) \quad (s, s^\prime =\pm),  \nonumber\\
&& F_{{\bf k},{\bf q}}^{s s^\prime} = -T\sum_{\varepsilon_n} {\cal G}^{s}_{\varepsilon_n}({\bf k}) {\cal G}^{s^\prime}_{\varepsilon_n}({\bf k}+{\bf q}) = - \frac{f( s E^s_{{\bf k}+{\bf q}} )-f(s^\prime E^{s^\prime}_{\bf k} )}{s E^s_{{\bf k}+{\bf q}} -s^\prime E^{s^\prime}_{\bf k} }, \nonumber
\end{eqnarray}
and $f(x)=\frac{1}{e^{x/T}+1}$. The amplitude of the CDW gap $|W_{{\bf Q}_c}|$ is determined from ${\cal F}_{\rm CDW}$. 
Below, we will discuss the roles of various terms appearing in Eq. (\ref{eq:RKKY}).

\begin{figure}[t]
\includegraphics[width=\columnwidth]{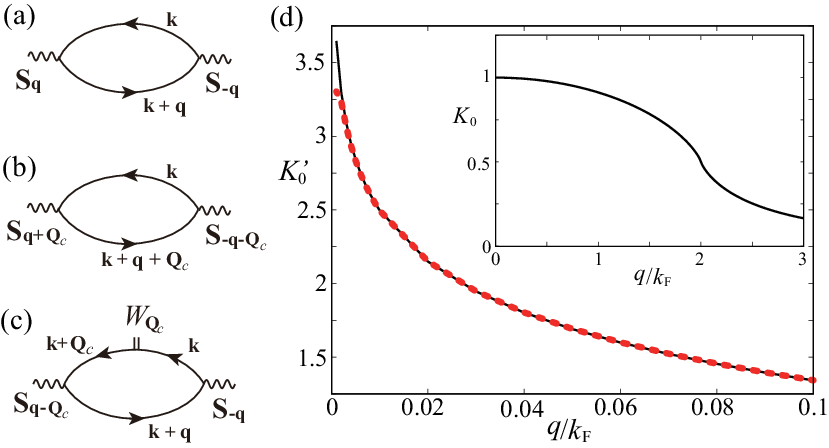}
\caption{(a)-(c) Feynman diagrams describing (a) the conventional RKKY interaction $K_0$, (b) the one of nesting origin $K_0^\prime$, and (c) the leading-order contribution $K_1$ from the CDW order $W_{{\bf Q}_c}$, where solid and wavy lines represent the normal Green's function and the localized spin ${\bf S}_{\bf q}$, respectively. (d) The $q$ dependences of $K_0^\prime$ (main panel) and $K_0$ (inset) at $T=0$ in the normal state without the CDW order, where in the main panel, black solid and red dotted curves correspond to the cases of $\delta=0$ and $\delta/E_F=0.002$, respectively. Both $K_0$ and $K_0^\prime$ are normalized by $N(0)$. \label{fig:fig1}}
\end{figure}

In the normal state without the CDW order, i.e., in the limit of $|W_{{\bf Q}_c}| \rightarrow 0$, we have $E^+_{\bf k}\rightarrow \xi_{\bf k}$, $E^-_{\bf k}\rightarrow -\xi_{{\bf k}+{\bf Q}_c}$, $g_{\bf k}\rightarrow 1$, and $f_{\bf k} \rightarrow 0$, so that the coefficients are reduced to
\begin{eqnarray}\label{eq:normal}
K_0 &=&  \frac{1}{2}\sum_{\bf k}  \big( F^{++}_{{\bf k}, {\bf q}} + F^{--}_{{\bf k}, {\bf q}} \big)\big|_{|W_{{\bf Q}_c}| \rightarrow 0} =  - \sum_{\bf k}  \frac{f(\xi_{{\bf k}+{\bf q}}) -f(\xi_{\bf k})}{\xi_{{\bf k}+{\bf q}}-\xi_{\bf k}}, \nonumber\\
K_0^\prime &=& \frac{1}{2}\sum_{\bf k} F^{-+}_{{\bf k}, {\bf q}}\big|_{|W_{{\bf Q}_c}| \rightarrow 0} =- \frac{1}{2}\sum_{\bf k} \frac{f(\xi_{{\bf k}+{\bf Q}_c+{\bf q}}) -f(\xi_{\bf k})}{\xi_{{\bf k}+{\bf Q}_c+{\bf q}}-\xi_{\bf k}} ,\nonumber\\
K_1&=&K_2=0  \qquad (|W_{{\bf Q}_c}| \rightarrow 0) .
\end{eqnarray} 
$K_0$ and $K_0^\prime$ terms correspond to the Feynman diagrams shown in Fig. \ref{fig:fig1} (a) and (b), respectively. These terms can be calculated for a specific Fermi surface satisfying the nesting condition 
\begin{equation}\label{eq:nesting}
\xi_{{\bf k}+{\bf Q}_c} = -\xi_{\bf k} + \delta .
\end{equation}
To grasp the meanings of $K_0$ and $K_0^\prime$, we first consider a simplified model in which it is assumed that $\delta$ measuring the deviation from the perfect nesting is a constant \cite{incomme_Rice_prb_70, FeSC_Vorontsov_prb_09, AFSC_Ikeda_prb_10, AFSC_Aoyama_prb_11, AFSC_Aperis_prl_10}, namely, the ${\bf k}$ dependence of $\xi_{\bf k}$ except Eq. (\ref{eq:nesting}) is basically irrelevant. With the use of Eq. (\ref{eq:nesting}), $K_0$ and $K_0^\prime$ in Eq. (\ref{eq:normal}) can be written as a function only of $\xi_{\bf k}$. At this point, the nesting effect has already been incorporated via Eq. (\ref{eq:nesting}), so that the problem is reduced to perform the integration over $\xi_{\bf k}$. As an analytically tractable $\xi_{\bf k}$, we use the ideally isotropic 3D Fermi surface as in the conventional RKKY theory. Then, $K_0$ at $T=0$ can be evaluated as $K_0({\bf q})=\chi \big(\frac{q}{k_F}\big)$ with 
\begin{equation}
\chi(x)= \frac{N(0)}{2 x} \Big[ x + \Big\{1- \big(\frac{x}{2} \big)^2 \Big\}  \ln \Big| \frac{1+\frac{x}{2}}{1-\frac{x}{2}}\Big| \Big] \nonumber\
\end{equation}
and density of states on the Fermi surface $N(0)$. As shown in the inset of Fig. \ref{fig:fig1} (d), $K_0$ does not have a peak, so that a characteristic wave vector does not exist, which together with an anomaly at $q=2k_F$, results in an oscillating power-law decay in its real-space representation. Thus, $-\sum_{\bf q}K_0({\bf q}) \, {\bf S}_{\bf q}\cdot{\bf S}_{-{\bf q}}$ term in Eq. (\ref{eq:RKKY}) corresponds to the conventional RKKY interaction.

In the same manner, $K_0^\prime$ at $T=0$ can be evaluated as $K_0^\prime ({\bf q}) =\chi^\prime \big(\frac{q}{k_F}\big)$ with 
\begin{equation}
\chi^\prime(x) =  \frac{N(0)}{2x} \int_0^1  d k \, k \ln \Big| \frac{k^2-kx-1+\frac{1}{2}(x^2-\frac{\delta}{E_F})}{k^2+k x -1+ \frac{1}{2}(x^2-\frac{\delta}{E_F})}\Big|. \nonumber 
\end{equation} 
As shown in the main panel of Fig. \ref{fig:fig1} (d), $K_0^\prime$ diverges toward $q=0$ in the perfect nesting case of $\delta =0$ (see the solid curve), and the divergence is suppressed by a nominal imperfection of the nesting, i.e., nonzero $\delta$ (see the red dotted curve). Since $K_0^\prime ({\bf q})$ becomes significantly large near ${\bf q}=0$, $-\sum_{\bf q} K_0^\prime ({\bf q}) \, S_{{\bf q}+{\bf Q}_c}\cdot{\bf S}_{-{\bf q}-{\bf Q}_c}$ term in Eq. (\ref{eq:RKKY}) favors the ${\bf S}_{\pm {\bf Q}_c}$ component, suggesting that spin correlation tends to develop at ${\bf Q}_c$. This is qualitatively consistent with results in Refs. \cite{RKKY-nesting_hayami_prb_17, Gd2PdSi3_RKKY-nesting_Dong_prl_24} where the association between the nesting and the RKKY interaction is discussed. We note that whether ${\bf Q}_c$ is actually selected or not as the ordering vector of ${\bf S}_i$ depends on not only the imperfection of the nesting $\delta$ but also the structure of the lattice on which ${\bf S}_i$'s are sitting, particularly for the classical localized spin ${\bf S}_i$ having an almost fixed length on the discrete lattice. 

Now, we shall turn to the CDW phase where ${\cal F}_{\rm CDW}$ can be expressed in the form expanded with respect to the CDW order parameter $W_{{\bf Q}_c}$ as ${\cal F}_{\rm CDW}= \alpha |W_{{\bf Q}_c}|^2 + \beta |W_{{\bf Q}_c}|^4$ with (for details, see Ref. \cite{supple})
\begin{eqnarray}\label{eq:coeff}
\alpha &=&  \frac{1}{|g|}  +  T\sum_{\varepsilon_n, {\bf k}}  {\cal G}_{\varepsilon_n}({\bf k}) {\cal G}_{\varepsilon_n}({\bf k}+{\bf Q}_c) \nonumber\\
&=& N(0)\Big[ \ln\frac{T}{T_{c0}}+2\pi T\sum_{\varepsilon_n}\Big( \frac{1}{2|\varepsilon_n|} - \frac{1}{2|\varepsilon_n|+is_{\varepsilon_n}\delta }\Big)\Big] , \nonumber\\
\beta &=& \frac{T}{2}\sum_{\varepsilon_n, {\bf k}} {\cal G}_{\varepsilon_n}({\bf k})^2 {\cal G}_{\varepsilon_n}({\bf k}+{\bf Q}_c)^2=\sum_{\varepsilon_n}\frac{2\pi T N(0)}{(2|\varepsilon_n|+is_{\varepsilon_n}\delta)^3}. 
\end{eqnarray} 
Here, $s_{\varepsilon_n}$ denotes the sign of $\varepsilon_n$ and the replacement $\sum_{\bf k} = N(0)\int_{-\infty}^{\infty} d\xi_{\bf k}$ has been used assuming that the the energy scale of the CDW is much smaller than $E_F$, i.e., the same approximation as that for the weak-coupling BCS theory in the context of superconductivity. The CDW transition temperature for $\delta=0$ is denoted by $T_{c0}$, whereas the one for $\delta \neq 0$ is determined by the condition $\alpha=0$. Since for the parameters used here, the second-order CDW transition occurs as in many of the CDW-hosting magnets introduced earlier \cite{Er5Ir4Si10_Galli_prl_00, RNiC2_Shimomura_prb_16, EuAl4_Shimomura_jpsj_19, RTe3_CDWtransition_Ru_prb08}, the CDW gap is given by $|W_{{\bf Q}_c}|^2 = - \alpha/(2\beta)$. 
  
To discuss how the spin-spin interaction is modified by the CDW order $W_{{\bf Q}_c}$, we first consider the leading-order contribution in the expansion with respect to $W_{{\bf Q}_c}$. Since $g_{\bf k}=1+{\cal O}(|W_{{\bf Q}_c}|^2)$, $f_{\bf k}= W_{{\bf Q}_c}/\gamma_{\bf k}+{\cal O}(|W_{{\bf Q}_c}|^3)$, and $F^{ss^\prime}_{{\bf k},{\bf q}}=F^{ss^\prime}_{{\bf k},{\bf q}}|_{W_{{\bf Q}_c}=0}+{\cal O}(|W_{{\bf Q}_c}|^2)$, it turns out that $K_1$ term in Eq. (\ref{eq:RKKY}) is the lowest-order contribution taking the form of
\begin{equation}\label{eq:1st}
- W_{{\bf Q}_c} \, \sum_{\bf q}\tilde{K}_1({\bf q})  {\bf S}_{{\bf q}-{\bf Q}_c}\cdot {\bf S}_{-{\bf q}} - W^\ast_{{\bf Q}_c} \, \sum_{\bf q} \tilde{K}_1(-{\bf q})  {\bf S}_{{\bf q}+{\bf Q}_c}\cdot {\bf S}_{-{\bf q}}
\end{equation} 
with   
\begin{eqnarray}
&& \tilde{K}_1({\bf q}) = -T\sum_{\varepsilon_n}\sum_{\bf k}  {\cal G}_{\varepsilon_n}({\bf k})  {\cal G}_{\varepsilon_n}({\bf k}+{\bf Q}_c) \nonumber\\
&& \qquad \times \Big[ {\cal G}_{\varepsilon_n}({\bf k}+{\bf q})+{\cal G}_{\varepsilon_n}({\bf k}+{\bf Q}_c-{\bf q})\Big] \\
&&= -4 N(0) \pi T\sum_{\varepsilon_n}\Big\langle \frac{ 2  |\varepsilon_n| \, ({\bf v}_{\bf k}\cdot{\bf q} +2 \delta ) }{ \big[4\varepsilon_n^2 + \delta^2 \big] \big[ 4\varepsilon_n^2+({\bf v}_{\bf k}\cdot {\bf q}+\delta)^2 \big] } \Big\rangle_{\rm FS}, \nonumber
\end{eqnarray}
where $\langle \rangle_{\rm FS}$ denotes the angle average on the Fermi surface. Note that this $K_1$ term is first order in $W_{{\bf Q}_c}$. The associated Feynman diagram is shown in Fig. \ref{fig:fig1} (c). Below, we shall take a closer look at the role of the $K_1$ term in Eq. (\ref{eq:1st}). 

Suppose that ${\bf q}_m$ is a candidate ordering vector of ${\bf S}_i$ expected from the normal-state RKKY interactions $K_0$ and $K_0^\prime$. Then, Eq. (\ref{eq:1st}) means that additional modes of ${\bf q}_m\pm{\bf Q}_c$ can be induced by the CDW order $W_{{\bf Q}_c}$. This can be interpreted in a different manner as follows: when several candidate ${\bf q}_m$'s are energetically almost degenerate as a result of the competition between $K_0$ and $K_0^\prime$ for $\delta \neq 0$, the CDW order may lift the degeneracy via the $K_1$ term in Eq. (\ref{eq:1st}).  
The emergence of this mode-coupling between ${\bf q}_m$ and ${\bf q}_m\pm{\bf Q}_c$ is the main result of this work. A similar mode-coupled fluctuation has been discussed in the different context of ferromagnetic superconductors \cite{mode_Watanabe_jpsj02}. Also, the inverse effect, namely, the spin-ordering effect on the electron density modulation, has been discussed in the context of skyrmion crystals \cite{CDW-RKKY_hayami_prb_21}. Since Eq. (\ref{eq:1st}) involves not the amplitude $|W_{{\bf Q}_c}|$ but the bare CDW order parameter $W_{{\bf Q}_c}=|W_{{\bf Q}_c}|e^{i\phi_c}$, the phase $\phi_c$ is relevant to the spin correlation. Since in the incommensurate case without impurity pinnings assumed here, a phase shift does not change the condensation energy, $\phi_c$ could be chosen such that the overall sign of Eq. (\ref{eq:1st}) be negative. It should also be noted that in the perfect nesting case of $\delta =0$, the mode-coupling term (\ref{eq:1st}) vanishes after the angle average on the Fermi surface, so that the imperfection of the nesting is important for the coupling between ${\bf q}_m$ and ${\bf Q}_c$. 

Concerning higher-order contributions in $|W_{{\bf Q}_c}|$, instead of picking them up, we will numerically evaluate the coefficients in Eq. (\ref{eq:RKKY}) without using the expansion.
As our interest here is in the effect of the CDW order, we introduce $\delta K_0 \equiv K_0-K_0|_{W_{{\bf Q}_c}=0}$ and $\delta K^\prime_0 \equiv K_0^\prime-K^\prime_0|_{W_{{\bf Q}_c}=0}$, the deviations from the normal-state values. In contrast to $\delta K_0$ and $\delta K_0^\prime$, $K_1$ and $K_2$ depend on not only the amplitude $|W_{{\bf Q}_c}|$ but also the CDW phase $\phi_c$. Since $\phi_c$ cannot be determined unless the lattice for ${\bf S}_i$ is specified, in the evaluation of $K_1$ and $K_2$, we simply set $\phi_c=0$. Also, we take the angle average $\langle \rangle_{\rm FS}$ on the spherical surface as a prototypical Fermi surface, so that all the coefficients depend only on $q=|{\bf q}|$.

\begin{figure}[t]
\includegraphics[scale = 0.5]{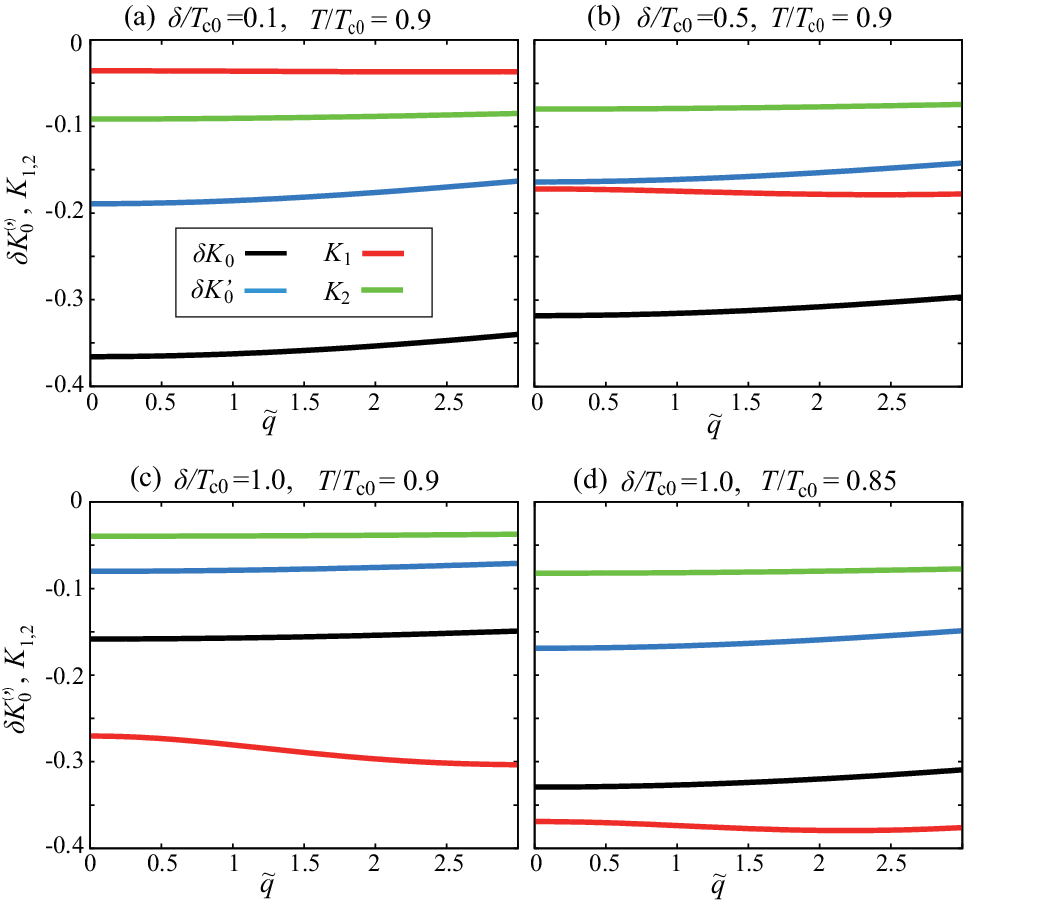}
\caption{The $q$ dependences of $\delta K_0$ (black), $\delta K_0^\prime $ (blue),  $K_1$ (red), and $K_2$ (green) obtained in the weak-coupling theory for (a) $\delta/T_{c0}=0.1$, (b) 0.5, and (c) 1.0 at the fixed temperature $T/T_{c0}=0.9$, and (d) $\delta/T_{c0}=1.0$ and $T/T_{c0}=0.85$, where $\tilde{q}=q \, (v_F/T_{c0})$ and $\delta K_0^{(\prime)}$ and $K_{1,2}$ are normalized by $N(0)$. \label{fig:fig2}}
\end{figure}
 
Figures \ref{fig:fig2} (a)-(c) show the $q$-dependences of $\delta K_0$, $\delta K^\prime_0$, $K_1$, and $K_2$ for different values of $\delta/T_{c0}$ at the fixed temperature $T/T_{c0}=0.9$ slightly below the CDW transition temperature, where the CDW gaps in (a)-(c) are calculated as $|W_{{\bf Q}_c}|/T_{c0}=0.89$, 0.84, and 0.62, respectively. Note that in the present weak-coupling theory where a characteristic energy scale is of the order of $T_{c0}$ much smaller than $E_F$, $q$ and $\delta$ are, respectively, scaled by $T_{c0}/v_F$ and $T_{c0}$ instead of $k_F$ and $E_F$. As one can see from Figs. \ref{fig:fig2} (a)-(c), both $\delta K_0$ and $\delta K^\prime_0$ take negative values, while the normal-state counter parts take positive values [see Fig. \ref{fig:fig1} (d)], which suggests that the CDW order weakens the normal-state RKKY interaction (for its real-space consequence, see Ref. \cite{supple}). Actually, as $\delta$ increases ($|W_{{\bf Q}_c}|$ decreases), $|\delta K_0|$ and $|\delta K^\prime_0|$ get smaller. Figure \ref{fig:fig2} (d) shows the lower-temperature result of Fig. \ref{fig:fig2} (c), where at this temperature $T/T_{c0}=0.85$, we have $|W_{{\bf Q}_c}|/T_{c0}=0.87$. Since $|W_{{\bf Q}_c}|$ becomes larger with decreasing temperature, the suppression of the RKKY interaction becomes more remarkable at lower temperatures [compare Figs. \ref{fig:fig2} (c) and (d)]. Such a tendency is also the case for $K_2$ (see green curves in Fig. \ref{fig:fig2}). Concerning $K_1$ represented by red curves in Fig. \ref{fig:fig2}, with increasing $\delta$ [see Figs. \ref{fig:fig2} (a)-(c)] or decreasing temperature [compare Figs. \ref{fig:fig2} (c) and (d)], $|K_1|$ becomes larger, namely, the mode coupling between ${\bf q}$ and ${\bf q}\pm{\bf Q}_c$ can become stronger. Note that the overall sign of the $K_1$ term depends on the CDW phase $\phi_c$.   

\begin{figure}[t]
\includegraphics[scale = 0.45]{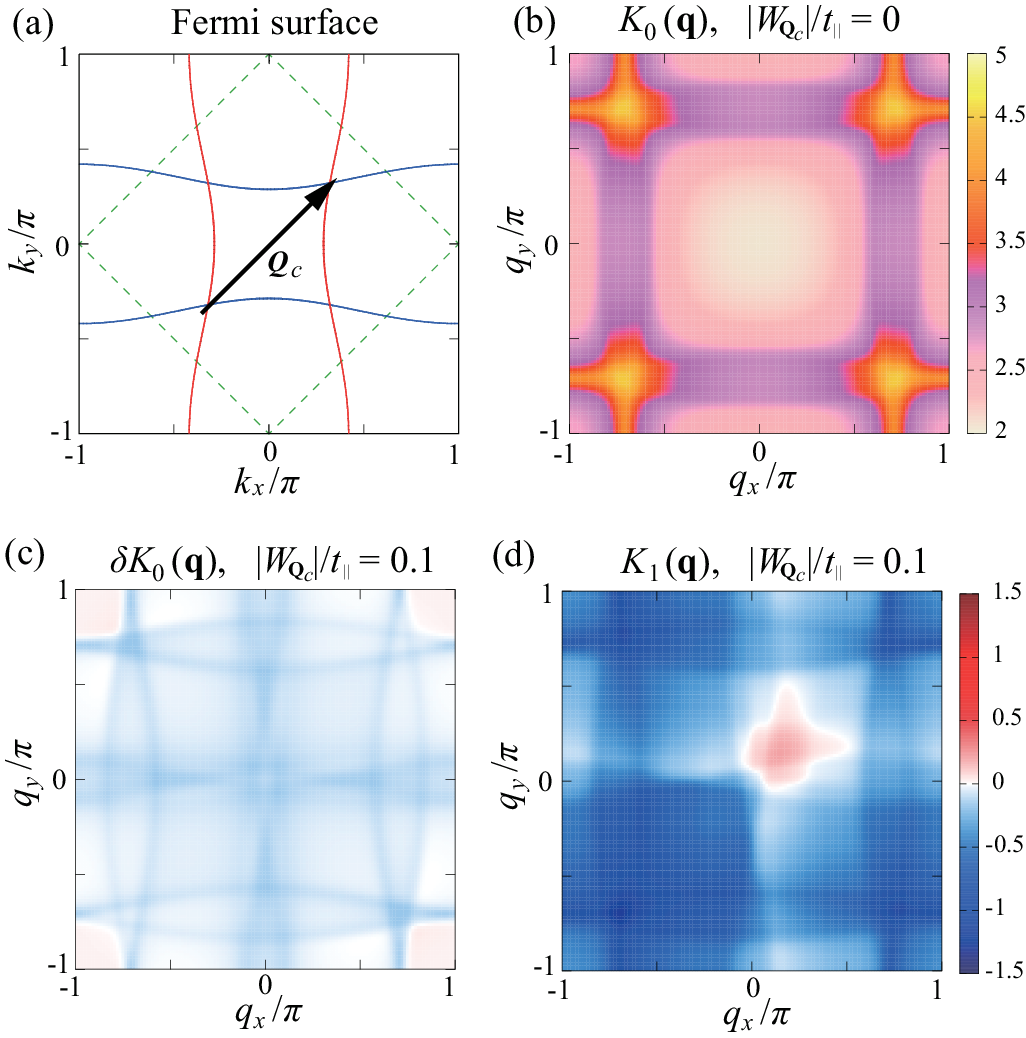}
\caption{Results obtained for the tight-binding model (\ref{eq:tight}) with the parameters associated with $R$Te$_3$, $t_\perp/t_\parallel = 0.185$, $\mu/t_\parallel = -2\cos(k_F)$, and $k_F = \frac{5\pi}{14}$. (a) The Fermi surface consisting of the $p_x$- (red curve) and $p_y$-orbitals (blue curve), where an arrow denotes the nesting vector ${\bf Q}_c=2k_F(1,1)$ and a green diamond represents a Brillouin zone for the larger unit cell (see the text). (b)-(d) The ${\bf q}$ dependences of (a) $K_0$ in the normal state, and (c) $\delta K_0$ and (d) $K_1$ in the CDW state with $|W_{{\bf Q}_c}|/t_\parallel =0.1$, where the intensity represented by the color bar is normalized by $N(0)$. \label{fig:fig3}}
\end{figure}

So far, we have discussed general aspects of the spin ordering in the CDW phase based on the weak-coupling theory taking account of the simplified nesting condition (\ref{eq:nesting}). Now, we apply the above discussion to the quasi-2D material $R$Te$_3$ where the interaction between the localized spins at $R$ sites is mediated by the conduction electrons and the features of the electronic structure can be captured by the following dispersion obtained from a tight-binding model on the Te-plane square lattice \cite{CeTe3_Brouet_prl_04, CDW_Kivelson_prb_06} 
\begin{equation}\label{eq:tight}
\xi_{{\bf k},1(2)} = -2t_\parallel \cos(k_{x(y)}a_0)+2t_\perp \cos(k_{y(x)}a_0)-\mu
\end{equation}
with $t_\parallel:t_\perp  = 2.0:0.37$, the Te-Te spacing $a_0$, and the band index 1 and 2 representing the $p_x$ and $p_y$ orbitals, respectively. In the model, we consider the single Te plane, taking the unit cell of the square lattice of lattice constant $a_0$, although in the real 3D materials having out-of-plane interactions, a basal plane of the 3D unit cell is rotated by 45 $^\circ$ compared with the square unit and has the larger side length of $a=\sqrt{2} a_0$ \cite{CeTe3_Brouet_prl_04, CDW_Fang_prl_07, CDW_Mazin_prb_08}. Referring to the association between ${\bf Q}_c$ and $\mu$ in this model \cite{CDW_Kivelson_prb_06}, we set $\mu = -2 t_\parallel \cos(k_F a_0)$ with $k_F=\frac{5}{14} \frac{\pi}{a_0}$ such that ${\bf Q}_c$ at the low temperature $T/t_\parallel = 0.04$ be close to the experimental value of ${\bf Q}_c\approx \frac{5}{7}\frac{\pi}{a_0}(1,1)$ \cite{CeTe3_Brouet_prl_04}. Note that $|{\bf Q}_c|=\frac{5}{7}\frac{\pi}{a_0}\sqrt{2} = \frac{5}{7} \times \frac{2\pi}{a}$ corresponds to $\frac{2}{7} \times \frac{2\pi}{a}$ for the rotated and larger unit cell \cite{CDW_Fang_prl_07}. For the obtained Fermi surface shown in Fig. \ref{fig:fig3} (a), we calculate the coefficients of the RKKY interaction (\ref{eq:RKKY}) in the cases without and with the CDW gap $|W_{{\bf Q}_c}|$. The extension of our analysis to the two-band case of Eq. (\ref{eq:tight}) is straightforward; we simply sum up the contributions from the $p_x$ and $p_y$ orbitals as they are decoupled in the associated model Hamiltonian \cite{CDW_Kivelson_prb_06} describing the single-${\bf Q}_c$ CDW state observed in $R$Te$_3$

Figure \ref{fig:fig3} (b) shows the ${\bf q}$ dependence of $K_0$ in the normal state of $|W_{{\bf Q}_c}|=0$. $K_0$ exhibits broad peaks at ${\bf q}= \frac{5}{7}\frac{\pi}{a_0} (\pm 1, \pm 1)$, suggesting that the spin correlation associated with the nesting vector tends to develop. Since in the present lattice model, the $K_0^\prime$ term in Eq. (\ref{eq:RKKY}) is merely the ${\bf Q}_c$-shifted version of the $K_0$ term, only $K_0$ is shown Fig. \ref{fig:fig3}. Figures \ref{fig:fig3} (c) and (d) show the ${\bf q}$ dependences of $\delta K_0$ and $K_1$ in the CDW state with $|W_{{\bf Q}_c}|/t_\parallel=0.1$ and $\phi_c =0$. The $K_2$ contribution (not shown) is negligibly small. Although $|W_{{\bf Q}_c}|$ should be determined from the associated ${\cal F}_{\rm CDW}$, it is given by hands as our interest here is in the difference between the cases without and with the CDW order. We have checked that the following results are qualitatively unchanged for different values of $|W_{{\bf Q}_c}|$. One can see from Figs. \ref{fig:fig3} (c) and (d), $\delta K_0$ basically takes a negative value, suppressing the normal-state RKKY interaction, and $K_1$ is nonzero, yielding the mode-coupling between $S_{{\bf q}\pm{\bf Q}_c}$ and $S_{-{\bf q}}$, which is consistent with the weak-coupling results shown in Fig. \ref{fig:fig2}. Since near ${\bf q}=0$, $|K_1|$ is larger in the ${\bf Q}_c$ direction parallel to the $(1,1)$ direction than in the orthogonal $(-1,1)$ direction, the magnetic wave vector parallel to ${\bf Q}_c$ may become more favorable (although $K_1$ shows a sign change along the $(1,1)$ direction, the net sign of the $K_1$ term depends on the CDW phase $\phi_c$). In that case, the spin correlation would develop at $q_m$ and $q_m \pm \frac{2}{7}$ in units of $\frac{2\pi}{a}$ for the rotated and larger unit cell.

In experiments, the CDW transition and a successive lower-temperature magnetic transition have been observed in several metallic systems \cite{RNiC2_Shimomura_prb_16, RNiC2_Hanasaki_prb_17, EuAl4_Shimomura_jpsj_19, EuAl4_Kaneko_jpsj_21, EuAl4_Takagki_natcom_22, EuAl4_Gen_prb_24, CeTe3_Okuma_screp_20, GdTe3_Raghavan_npjq_24, TbTe3_Pfuner_jpcm_12, TbTe3_Chillal_prb_20, RTe3_Higashihara_prb_24, DyTe3_Akatsuka_natcom_24} among which the $R$Te$_3$ families are more relevant to the present theoretical work since it is well established that the Fermi surface nesting is essential for the CDW order and the CDW ordering vector ${\bf Q}_c$ is incommensurate \cite{CeTe3_Brouet_prl_04, CDW_Fang_prl_07, CDW_Mazin_prb_08} as assumed here (in EuAl$_4$, the origin of the CDW seems to remain unknown at present \cite{EuAl4_Kaneko_jpsj_21}, and in Er$_5$Ir$_4$Si$_{10}$ and (Gd,Tb)NiC$_2$, ${\bf Q}_c$ is commensurate near the magnetic transition temperature \cite{Er5Ir4Si10_Galli_prl_00, RNiC2_Shimomura_prb_16}).  
In TbTe$_3$, it has been reported that the lowest-temperature magnetic order is characterized by the Bragg peak at ${\bf q}_{m}+{\bf Q}_c$ with the CDW ordering vector ${\bf Q}_c$ and a higher-temperature magnetic ordering vector ${\bf q}_m$ \cite{TbTe3_Chillal_prb_20}, and such a coupling between ${\bf q}_m$ and ${\bf Q}_c$ has also been observed in DyTe$_3$ \cite{DyTe3_Akatsuka_natcom_24}, although ${\bf q}_m$'s differ in two cases. Our result on the mode-coupled $K_1$ term is qualitatively consistent with the observation. Also, a similar coupling term has been phenomenologically introduced to explain a magnetic order in GdTe$_3$ \cite{GdTe3_Raghavan_npjq_24}. 

On the other hand, in GdTe$_3$, the magnetic-dipole and superexchange interactions might be more relevant \cite{GdTe3_interaction_Liu_prm_20}. In addition, the Kondo coupling can induce the Kondo effect rather than the RKKY interaction, as is suggested in CeTe$_3$ \cite{CeTe3_Kondo_Ru_prb_06}. Possibly due to these competitions,
the magnetic ordering vectors some of which have out-of-plane components depend on specific materials, but we believe that the CDW induced mode-coupling presented here would play an important role for the magnetic properties of this class of magnets. 
What kind of spin structure is actually favored by the interaction (\ref{eq:RKKY}) on a specific lattice would be an important question. 
Noting that the CDW phase $\phi_c$ is associated with the CDW dynamics \cite{C0018, C0070, C0014, Et-Shapiro_Thorne_prb_87, Matsukawa_JJAP_1987, FA_shapiro,FA_frqmix, RTe3_sliding_Sinchenko_prb_12} and it directly enters the mode coupling, how the external-field-driven CDW sliding dynamics affects the magnetism via $\phi_c$ also turns out to be an open question \cite{DyTe3_Akatsuka_natcom_24}.
We will leave these issues for our future work, together with the extension to a commensurate-${\bf Q}_c$ case.

\begin{acknowledgments}
The author thanks Y. Niimi, Y. Okada, and R. Okuma for useful discussions. This work is partially supported by JSPS KAKENHI Grant No. JP21K03469, JP23H00257, and JP24K00572.
\end{acknowledgments}

\hspace{2cm}
\pagebreak

\onecolumngrid
\hspace{2cm}
\begin{center}
\textbf{\large Supplemental Material for ''RKKY interaction in the presence of a charge-density-wave order''}\\[.2cm]
\end{center}

\twocolumngrid
\section{Mean-field CDW Hamiltonian}
We derive the mean-field CDW Hamiltonian (1) in the main text. Following the theoretical work by Balseiro and Falikov \cite{BF_CDW-theory_79}, we start from the phonon part of the Hamiltonian ${\cal H}_p$ which is given by 
\begin{equation}\label{eq:phonon}
{\cal H}_p = \sum_{\bf q} \omega_{\bf q} \hat{b}^\dagger_{\bf q} \hat{b}_{\bf q}  + i \sum_{{\bf k},{\bf q}, s} D_{\bf q} \hat{c}^\dagger_{{\bf k}+{\bf q},s} \hat{c}_{{\bf k},s}( \hat{b}_{\bf q}-\hat{b}^\dagger_{-{\bf q}}),
\end{equation}
where $\hat{b}^\dagger_{\bf q}$ and $\hat{b}_{\bf q}$ ($\hat{c}^\dagger_{{\bf k},s}$ and $\hat{c}_{{\bf k},s}$) denote creation and annihilation operators of the phonons (electrons), respectively. In Eq. (\ref{eq:phonon}), the first term describes the phonon energy with the dispersion $\omega_{\bf q}$, and the second term describes the electron-phonon coupling with coupling constant $D_{\bf q}$ which will be taken to be a constant $D$ for simplicity. By using a canonical transformation \cite{Canonical_Frohlich_prs_52}, we obtain the phonon-mediated effective electron-electron interaction 
\begin{equation}\label{eq:effective}
{\cal H}_{\rm eff} =  \sum_{\bf q}\sum_{{\bf k}, {\bf k}', s, s'} \, V_{{\bf k},{\bf k}'}^{{\bf q}} \, \hat{c}_{{\bf k}+{\bf q},s}^\dagger \hat{c}_{{\bf k},s} \hat{c}^\dagger_{{\bf k}'-{\bf q},s'}\hat{c}_{{\bf k}',s'}
\end{equation}
with $V_{{\bf k},{\bf k}'}^{\bf q}=\frac{D^2}{4}\Big( \frac{2\omega_{\bf q}^2}{(\varepsilon_{\bf k}-\varepsilon_{{\bf k}-{\bf q}})^2-\omega_{\bf q}^2} + \frac{2\omega_{\bf q}^2}{(\varepsilon_{{\bf k}'}-\varepsilon_{{\bf k}'+{\bf q}})^2-\omega_{\bf q}^2} \Big)=V_{{\bf k}^\prime,{\bf k}}^{-{\bf q}}$. 
As is well known, in Eq. (\ref{eq:effective}), an attractive interaction near ${\bf q}=0$ is relevant to the phonon-mediated $s$-wave superconductivity. In the case of the CDW characterized by the nesting vector ${\bf Q}_c$, an associated attractive interaction corresponds to the ${\bf q}={\bf Q}_c$ mode. Thus, we assume $V_{{\bf k},{\bf k}'}^{\bf q}=  - \frac{1}{8}|g| \big( \delta_{{\bf q},{\bf Q}_c}+\delta_{{\bf q},-{\bf Q}_c}) \, \Lambda_{\bf k}\Lambda_{{\bf k}+{\bf q}}\Lambda_{{\bf k}'} \Lambda_{{\bf k}'-{\bf q}}$ with an interaction strength $g$ for the ${\bf q}={\bf Q}_c$ mode. Since the electrons near the Fermi surface is relevant to the CDW instability, we have introduced $\Lambda_{\bf k}$ which takes 1 only for $|\xi_{\bf k}|<\omega_c$ with a cutoff energy $\omega_c$ \cite{BF_CDW-theory_79}. 
 
Now, we introduce the CDW order parameter
\begin{equation}
W_{{\bf Q}_c} = -\frac{|g|}{2} \sum_{{\bf k}',s} \Lambda_{{\bf k}'} \Lambda_{{\bf k}'-{\bf Q}_c} \langle \hat{c}^\dagger_{{\bf k}'-{\bf Q}_c,s}\hat{c}_{{\bf k}',s} \rangle, \nonumber\\
\end{equation}
where $\langle \rangle$ denotes the thermal average. Then, $W^\ast_{{\bf Q}_c} = W_{-{\bf Q}_c}$ is satisfied. The use of the mean-field approximation yields
\begin{eqnarray}
 {\cal H}_{\rm eff} &\simeq& \frac{1}{2}\sum_{{\bf k},s} \Big( W_{{\bf Q}_c} \, \Lambda_{\bf k}\Lambda_{{\bf k}+{\bf Q}_c} \hat{c}_{{\bf k}+{\bf Q}_c,s}^\dagger \hat{c}_{{\bf k},s}  \\
 && + W_{-{\bf Q}_c} \, \Lambda_{{\bf k}}\Lambda_{{\bf k}-{\bf Q}_c} \hat{c}_{{\bf k},s}^\dagger \hat{c}_{{\bf k}-{\bf Q}_c,s} \Big) + \frac{1}{|g|}|W_{{\bf Q}_c}|^2. \nonumber
\end{eqnarray}
When we take $\Lambda_{\bf k}=1$ for simplicity, we obtain the mean-field Hamiltonian (1) in the main text. It is known that the so-obtained mean-field Hamiltonian can describe the qualitative aspects of the CDW order in $R$Te$_3$ \cite{CDW_Kivelson_prb_06}. Although the approximation $\Lambda_{\bf k}=1$ indicates that in contrast to realistic situations, the CDW gap $|W_{{\bf Q}_c}|$ opens over the whole Fermi surface, it would work well as long as temperature is not far below the CDW transition $T_{\rm CDW}$, since near $T_{\rm CDW}$ where the gap $|W_{{\bf Q}_c}|$ is small, quasiparticles can be thermally activated even in the presence of the CDW gap and thus, the effect of the momentum dependence of $\Lambda_{\bf k}$ on the quasiparticle excitation should be relatively weak. In the main text, focusing on such a moderate-temperature region, we discuss the direct coupling between the CDW order $W_{{\bf Q}_c}$ and localized spins ${\bf S}_{\bf q}$. 

\section{Coefficients $\alpha$ and $\beta$ in ${\cal F}_{\rm CDW}$}
In the main text, the CDW part of the free energy is expressed as ${\cal F}_{\rm CDW}= \alpha |W_{{\bf Q}_c}|^2 + \beta |W_{{\bf Q}_c}|^4$. Here, we derive the concrete expressions of the coefficients $\alpha$ and $\beta$ by using the weak-coupling approximation, the same approximation as that for the weak-coupling BCS theory in the context of superconductivity, where the energy scale of the CDW is much smaller than $E_F$ and the summation over ${\bf k}$ can be replaced with $\sum_{\bf k}=N(0)\int_{-\infty}^\infty d \xi_{\bf k} \langle \rangle_{\rm FS}$. Since the normal-state Green's function is given by ${\cal G}_{\varepsilon_n}({\bf k})=(i\varepsilon_n-\xi_{\bf k})^{-1}$, the coefficient $\alpha$ can be calculated as    
\begin{eqnarray}\label{eq:2nd}
\alpha &=&  \frac{1}{|g|}  +  T\sum_{\varepsilon_n, {\bf k}}  {\cal G}_{\varepsilon_n}({\bf k}) {\cal G}_{\varepsilon_n}({\bf k}+{\bf Q}_c) \nonumber\\
&=&  \frac{1}{|g|}  +  T\sum_{\varepsilon_n} N(0)\int_{-\infty}^\infty d \xi_{\bf k} \frac{1}{i\varepsilon_n-\xi_{\bf k}}\frac{1}{i\varepsilon_n-\xi_{{\bf k}+{\bf Q}_c}} \nonumber\\
&=&  \frac{1}{|g|}  +  T\sum_{\varepsilon_n} N(0)\int_{-\infty}^\infty d \xi_{\bf k} \frac{1}{i\varepsilon_n-\xi_{\bf k}}\frac{1}{i\varepsilon_n+\xi_{\bf k} -\delta} \nonumber\\
&=&  \frac{1}{|g|}  +  T\sum_{\varepsilon_n} N(0)\frac{-2\pi i s_{\varepsilon_n}}{2i\varepsilon_n-\delta} \nonumber\\
&=&  \frac{1}{|g|}  -  2\pi N(0) T\sum_{\varepsilon_n} \frac{1}{2|\varepsilon_n| +i s_{\varepsilon_n} \delta} ,
\end{eqnarray}
where the nesting condition $\xi_{{\bf k}+{\bf Q}_c}=-\xi_{\bf k}+\delta$ [Eq. (5) in the main text] has been used and $s_{\varepsilon_n}$ denotes the sign of $\varepsilon_n$.  
Noting that when the frequency summation $\sum_{\varepsilon_n}$ is performed first, $T\sum_{\varepsilon_n, {\bf k}}  {\cal G}_{\varepsilon_n}({\bf k}) {\cal G}_{\varepsilon_n}({\bf k}+{\bf Q}_c)$ can also be written as
\begin{eqnarray}\label{eq:nesting_fnc}
 T\sum_{\varepsilon_n, {\bf k}}  {\cal G}_{\varepsilon_n}({\bf k}) {\cal G}_{\varepsilon_n}({\bf k}+{\bf Q}_c) = \sum_{\bf k}\frac{f(\xi_{\bf k})-f(\xi_{{\bf k}+{\bf Q}_c})}{\xi_{{\bf k}+{\bf Q}_c}-\xi_{\bf k}}.
\end{eqnarray}
we find that the second term in Eq. (\ref{eq:2nd}) corresponds to the nesting function or the charge density susceptibility whose peak or divergence signals the occurrence of the CDW ordering. 
Actually, in the perfect nesting case of $\delta=0$, by introducing the cutoff energy $\omega_c$, we have $4\pi TN(0) \sum_{{\varepsilon_n}> 0}^{\omega_c}\frac{1}{2\varepsilon_n} \simeq  N(0) \ln\big( \frac{\omega_c/2\pi}{T}\big)$, so that with decreasing temperature $T$, $\alpha \simeq \frac{1}{|g|}- N(0)\ln\big( \frac{\omega_c/2\pi}{T}\big)$ decreases from a positive value and eventually goes to zero at a certain temperature $T_{c0}$ which corresponds to the CDW transition temperature. The logarithmic divergence originating from the nesting condition $\xi_{{\bf k}+{\bf Q}_c}=-\xi_{\bf k}$ is associated with the enhancement of the charge density susceptibility. 
By expressing $1/|g|$ with $T_{c0}$, we obtain 
\begin{equation}
\alpha= N(0)\Big[ \ln\frac{T}{T_{c0}}+2\pi T\sum_{\varepsilon_n}\Big( \frac{1}{2|\varepsilon_n|} - \frac{1}{2|\varepsilon_n|+is_{\varepsilon_n}\delta }\Big)\Big] .\nonumber
\end{equation}

In the same manner, $\beta$ can be calculated as
\begin{eqnarray}
\beta &=& \frac{T}{2}\sum_{\varepsilon_n, {\bf k}} {\cal G}_{\varepsilon_n}({\bf k})^2 {\cal G}_{\varepsilon_n}({\bf k}+{\bf Q}_c)^2 \nonumber\\
&=& \frac{1}{2}N(0)T\sum_{\varepsilon_n} \int_{-\infty}^\infty d \xi_{\bf k} \Big( \frac{1}{i\varepsilon_n-\xi_{\bf k}} \Big)^2 \Big( \frac{1}{i\varepsilon_n+\xi_{\bf k} -\delta} \Big)^2 \nonumber\\
&=& \frac{1}{2}N(0)T\sum_{\varepsilon_n} (-4\pi i s_{\varepsilon_n})\Big( \frac{1}{2i\varepsilon_n -\delta} \Big)^3 \nonumber\\
&=& \sum_{\varepsilon_n}\frac{2\pi T N(0)}{(2|\varepsilon_n|+is_{\varepsilon_n}\delta)^3}, \nonumber
\end{eqnarray} 
where we have performed the complex integral with the use of the residue theorem for poles of order two. 

\section{Decay rate of the RKKY interaction in the CDW phase}
\begin{figure}[t]
\includegraphics[scale =0.55]{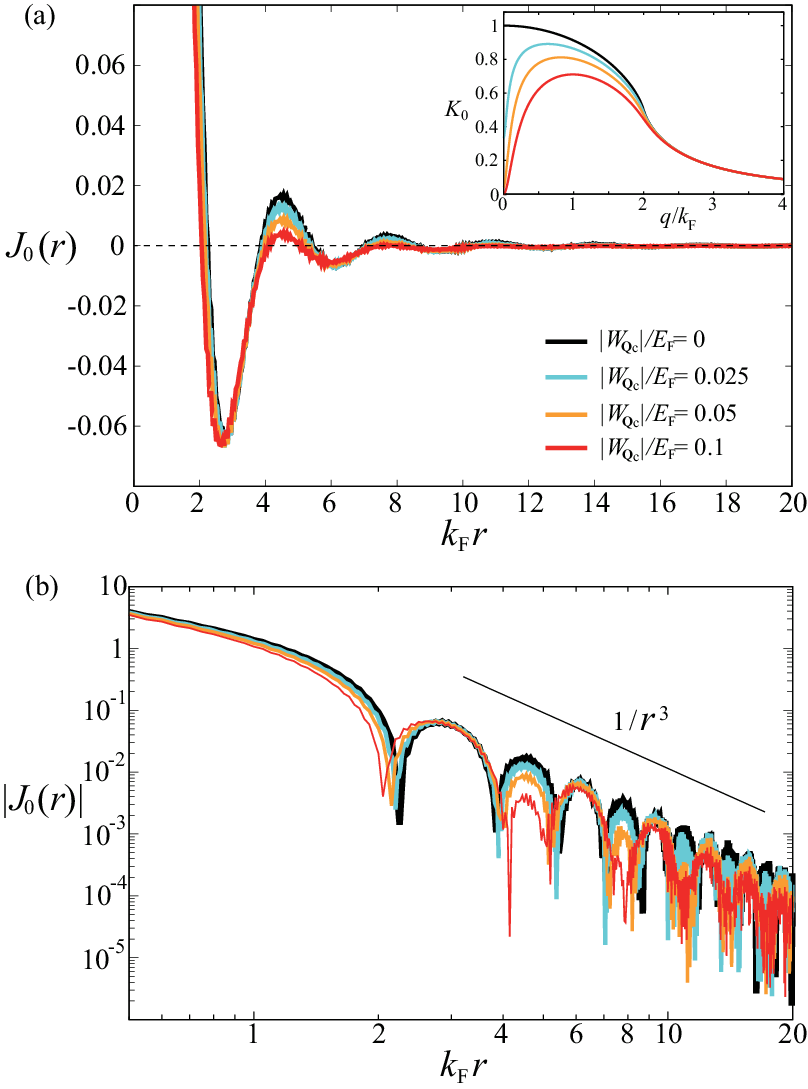}
\caption{The spatial dependence of the RKKY interaction obtained for $|W_{{\bf Q}_c}|/E_F=0$ (black), 0.025 (cyan), 0.05 (orange), and 0.1 (red) at $\delta/E_F=0.002$. (a) The regular plot of the RKKY interaction $J_0(r)$ and (b) the log-log plot of its absolute value $|J_0(r)|$, where $J_0(r)$ is normalized by $[N(0)]^2 E_F$. In (a), the inset shows the momentum dependence of $K_0$, where the notations are the same as those in the inset of Fig. 2 (d) in the main text. \label{fig:supple}}
\end{figure}

As discussed in the main text, the CDW order suppresses both the $K_0$ and $K_0^\prime$ terms the former of which corresponds to the conventional RKKY interaction showing the oscillating power-law decay. Since the latter $K_0^\prime(0)$ term originating from the Fermi surface nesting tends to favor the spin correlation near the ordering vector ${\bf Q}_c$, the $K_0$ term could be less relevant compared with the conventional case without the CDW order. Nevertheless, it is instructive to see how the presence of the CDW affects the decay rate of the $K_0$ term. 

To see how the CDW gap $|W_{{\bf Q}_c}|$ can modify the oscillating power-law decay of the RKKY interaction, we consider the following real-space representation of the $K_0$ term. 
\begin{eqnarray}\label{eq:RKKY}
J_0(r) &=& \sum_{\bf q} K_0({\bf q}) e^{i {\bf q} \cdot {\bf r}} \nonumber\\
&=& \frac{1}{(2\pi)^3} \int_0^\infty q^2 dq \int_0^{2\pi} d\phi \int_{-1}^{1} dx K_0(q) e^{iqr x} \nonumber\\
&=& \frac{1}{(2\pi)^2} \frac{1}{r}  \int_0^\infty dq \, qK_0(q) \sin (qr) .
\end{eqnarray}  
For comparison with the conventional RKKY interaction, we use the isotropic Fermi surface to calculate $K_0(q)$. Since our focus is on the effect of the CDW gap $|W_{{\bf Q}_c}|$ on $J_0(r)$, we will treat it as a free parameter here, although it should be determined from ${\cal F}_{\rm CDW}$.
  
Figure \ref{fig:supple} shows the spatial dependence of the RKKY interaction $J_0(r)$ defined by Eq. (\ref{eq:RKKY}). One can see from Fig. \ref{fig:supple} (a) that as the CDW gap $|W_{{\bf Q}_c}|$ develops, the long-wave-length part of $K_0(q)$ is gradually suppressed (see the inset) and resultantly, the amplitude of the oscillation becomes smaller (see the main panel). For the larger CDW gap (see the orange and red curves), the oscillating behavior in the long-length-scale region is hardly visible in this regular plot. Figure \ref{fig:supple} (b) shows the log-log plot of Fig. \ref{fig:supple} (a). One can see the conventional $r^{-3}$ decay of the RKKY interaction in the absence of the CDW gap (see the black curve). As the CDW gap develops, such a power-law behavior in the $3\leq k_F r \leq 10$ region is gradually modified, but it is still present in the longer length scale of $10 \leq  k_F r $. The decay rate is likely to be $r^{-3}$, i.e., the same as the one in the conventional RKKY interaction, although the interaction strength itself is suppressed by the CDW order.  

\end{document}